# The Design of Data Acquisition System for EAST Technical Diagnostic System

Ying Chen, Shi Li, Huazhong Wang, Yong Wang, and Bingjia Xiao

*Abstract*—EAST (Experimental Advanced Superconducting Tokamak) Technical Diagnostic System (TDS) is used to monitor the outlet temperature of all superconducting coils, in case of temperature anomaly, it will trigger safety interlock system to meet EAST device safety. The data acquisition system of TDS is in charge of continuous data acquisition of the nitrogen and helium temperature signals, TDS security alarm and long-term data storage. It supports continuous data acquisition and pulse data acquisition. The data acquisition of the nitrogen temperature signals is based on the PXI technology while obtaining the helium temperature signals from Lake Shore model 224 temperature monitors with VISA standard. After data conversion, all the data will be stored in MySQL and MDSPlus for long-term storage. It should output TDS fault signal and status signal to trigger the safety interlock system to take actions after threshold evaluation of key temperature signals. It publishes part of real-time TDS data to the cryogenic system and provides an information inquiry service to the TDS administrator. The system has been used in 2018 EAST campaign.

## I. Introduction

EAST (Experimental Advanced Superconducting Tokamak) is the world's first fully superconducting tokamak experimental device, and aims at steady-state long-pulse advanced high-performance operations [1]. The superconducting tokamak is different from the ordinary tokamak and it has special running style and higher risk [2]. In order to know the device running state in time and reduce the damage of the device, EAST engineering plant systems including the technical diagnostic system (TDS) are in charge of measuring key engineering parameters and outputting alarm signals to EAST safety interlock system. The EAST safety interlock system protects EAST device from potentially harmful operating under abnormal conditions [3, 4, 5]. During the EAST campaign, TDS should keep monitoring the outlet temperature of all superconducting coils, in case of temperature anomaly, it will trigger EAST safety interlock system to meet EAST device safety.

TDS has been upgraded after 2017 EAST campaign. This data acquisition system is designed for the upgraded TDS, abbreviated as TDS_DAQ. It is in charge of the data acquisition of temperature signals, TDS security alarm and long-term data storage. The following is the main requirements for TDS_DAQ.

(1) Continuous data acquisition. TDS mainly measures two kind of temperature signals: the nitrogen temperature signals and helium temperature signals, which are used to characterize the state of superconducting coils. An EAST campaign usually lasts several months, so TDS_DAQ should support 7 × 24h data acquisition within several months. Besides, for the EAST pulse experiments are shot by shot, TDS_DAQ should also support continuous long-pulse data acquisition.

(2) Multiple data storage. The original data storage structure should be maintained, so the TDS data will be stored with MySQL and MDSPlus [6]. MySQL stores real-time data of all engineering plant systems including TDS. In EAST, all experimental data including the TDS data should be stored with MDSPlus for long-term data storage.

(3) Security alarm. The outlet temperature can characterize the superconducting coils state, so TDS_DAQ should keep concerning on whether these temperatures are located in the normal range. After threshold evaluation of some key temperature signals, it should output TDS fault signal and status signal to trigger the safety interlock system to take actions.

(4) Data service. EAST cryogenic system, another engineering system, always needs some TDS data, so the TDS_DAQ should keep publishing TDS data to the cryogenic system. In addition, TDS_DAQ can provide some data service for the TDS administrator to easily get the alarm information and TDS data.

## II. System Architecture

The TDS has 236 nitrogen temperatures signals and 119 helium temperatures signals. The nitrogen temperature signals are all converted to voltage signals and can be acquired by the data acquisition card based on PXI. The helium temperature signals will be processed by Lake Shore model 224 temperature monitors [7]. Then these helium temperature signals can be read with VISA standard.

The TDS_DAQ is in charge of continuous data acquisition of the nitrogen and helium signals, data storage and TDS security

This work was supported by National Key R&D Program of China (Grant No: 2017YFE0300500, 2017YFE0300504).

Ying Chen is with the Institute of Plasma Physics, Chinese Academy of Sciences, Hefei, Anhui, 230031 China (e-mail: cheny@ipp.ac.cn).
Shi Li is with the Institute of Plasma Physics, Chinese Academy of Sciences, Hefei, Anhui, 230031 China (e-mail: lishi@ipp.ac.cn).
Huazhong Wang is with the Institute of Plasma Physics, Chinese Academy of Sciences, Hefei, Anhui, 230031 China (e-mail: hzwang@ipp.ac.cn).
Yong Wang is with the Institute of Plasma Physics, Chinese Academy of Sciences, Hefei, Anhui, 230031 China (e-mail: wayong@ipp.ac.cn).
Bingjia Xiao is with the Institute of Plasma Physics, Chinese Academy of Sciences, Hefei, Anhui, 230031 China (e-mail: bjxiao@ipp.ac.cn).

alarm. The system architecture is as shown in Fig.1. It can be divided into four modules: data communication module, data acquisition module, security alarm module, information inquiry module.

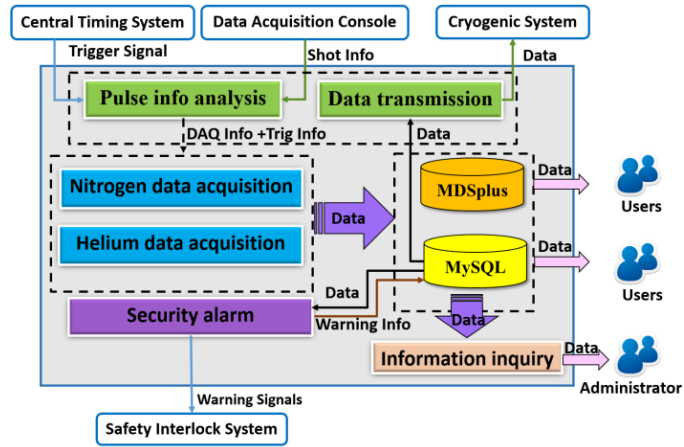

Fig. 1. System architecture

The data communication module is responsible for the data communication and data transmission with other systems. It includes pulse info analysis and data transmission. The pulse info analysis can get shot info from the data acquisition console via TCP/IP. The shot info describes the shot information such as current shot number and data acquisition time. Then the pulse info analysis waits for the trigger signal from the central timing system [8]. After get the shot info and trigger info, it starts up the data acquisition module to get the pulse data. The data transmission reads the real-time TDS data from the MySQL, and then publish the data to the cryogenic system.

The data acquisition module is responsible for data acquisition and storage of the nitrogen and helium data. For the different data acquisition mode, it can be divided into two parts: the nitrogen data acquisition and the helium data acquisition. The nitrogen data acquisition is based on PXI technology and provides 236 channels for nitrogen signals. After the processing of signal conditioners, the nitrogen temperature signals still need data conversion with linear formulas in the nitrogen data acquisition. The helium data acquisition is based on VISA. The 119 helium signals have been processed and converted by the Lake Shore model 224 temperature monitor. The helium data acquisition can obtain the helium signals from Lake Shore model 224 temperature monitors with the VISA standard, and doesn't need any more data conversion. The TDS data should be stored with two types: pulse data and continuous data. The data of one shot are called as pulse data while the 7×24h data seem as continuous data. The data acquisition module keeps acquiring the continuous data and stored into MySQL and MDSPlus. The data in MySQL will be constantly updated. After getting the shot info and trigger info from the data communication module, the data acquisition module start to get the pulse data and stored into MDSPlus for long-term storage. To distinguish the continuous data and the pulse data in MDSPlus, they are stored in MDSPlus day tree and MDSPlus pulse tree respectively.

The security alarm module monitors some key temperatures, in case of temperature anomaly, it shall trigger the safety interlock system to take actions. The security alarm module selects some temperature signals from different positions of superconducting coils as standard signals. It keeps threshold evaluation of these key temperature signals and outputs two safety signals to EAST safety interlock system: TDS status signal and TDS fault signal. The two signals will trigger different levels of protection, with the fault signal will cause a more serious result than the status signal.

The information inquiry module can provide some data services to the TDS administrator. The information inquiry module and the security alarm module help the understanding of the superconducting coils state under abnormal conditions. The TDS administrator uses the TDS information inquiry service to get all the alarm logs, draw the historical pulse data or continuous data from MDSPlus, and access the current real-time temperature data from MySQL. Experimenters also can use other tools such as EAST engineering data visualization website or WebScope [9] to view the real-time data or the historical data.

### III. SYSTEM IMPLEMENTATION

The TDS_DAQ is developed with LabVIEW. The following describes the implementation of TDS_DAQ.

#### A. Data Communication Module

The data communication module consists of two parts: pulse info analysis and data transmission. Fig.2 show the programs of the pulse info analysis.

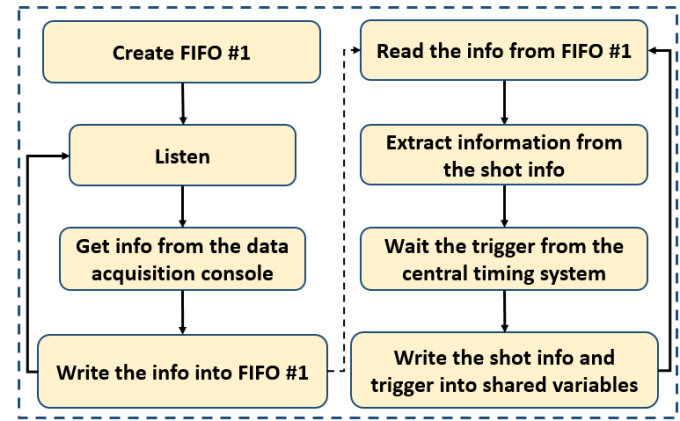

Fig. 2. Pulse info analysis

The pulse info analysis adopts the producer–consumer structure. The producer keeps listening to the data acquisition console. When a new shot is coming, the data acquisition console will send shot info to the data communication module. After getting the shot info, the pulse info analysis will write the info into FIFO #1 which has been created previously. The consumer keeps reading the shot info from FIFO #1, and extracts the key information such as shot number and data acquisition time. After the trigger is arriving, the shot info and trigger info will be wrote into shared variables which are shared with other devices in the same LAN and the data acquisition module will start to acquire the pulse data if the shared variables have been updated.

The cryogenic system is a critical plant engineering system which monitors and controls the cryogenic process and devices

[10, 11]. It should also keep running during the EAST campaign and always needs part of real-time TDS data. For this purpose, the data transmission should keep the connection with the cryogenic system and publish the real-time TDS data to the cryogenic system.

*B. Data Acquisition Module*

According to the different data acquisition methods, the data acquisition module has two parts: nitrogen data acquisition and helium data acquisition. The nitrogen data acquisition uses several NI PXI-6259 cards, while the helium data acquisition uses an industrial personal computer (IPC) to read the data from dozens of Lake Shore model 224 temperature monitors with VISA. The nitrogen data acquisition and helium data acquisition adopt the same software framework shown as Fig.3.

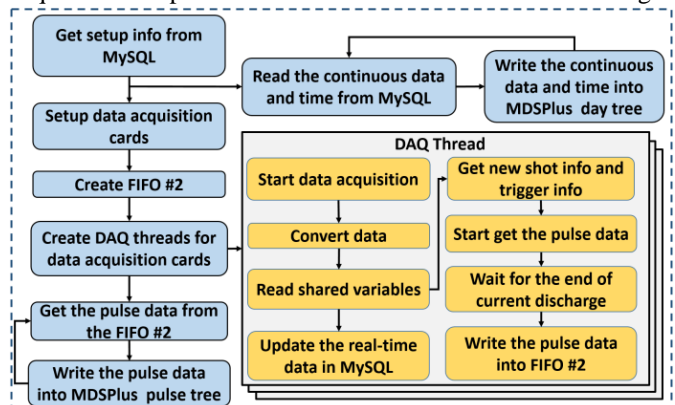

Fig. 3. Data acquisition

The data acquisition setup information and signals information are stored in MySQL. Firstly, the data acquisition program should get setup info from MySQL. After setup the data acquisition cards according to the setup info, it creates FIFO #2 and create DAQ threads for each data acquisition card. Then the main program will keep reading the pulse data from FIFO #2 while the DAQ threads write the pulse data into FIFO #2, and then writes the pulse data into the MDSPlus pulse tree remotely. Besides, the main program also keeps reading the continuous data and absolute time from MySQL, and then write into the MDSPlus day tree remotely. The MDSPlus day tree takes the date as the shot number.

In DAQ threads, each data acquisition card will start the continuous data acquisition. In the nitrogen data acquisition, the NI PXI-6259 data acquisition cards work in local and convert all data with linear formulas in each loop. In the helium data acquisition, each DAQ thread connects to the corresponding Lake Shore model 224 temperature monitor and don't need any more data conversion. After data conversion, each DAQ thread reads shared variables from the data communication module and check whether a new shot is coming, and updates the real-time data in MySQL. If a new shot is coming, it starts to get the pulse data and write the pulse data into FIFO #2 which will be accessed by the main program continuously after the current discharge is finished.

*C. Security Alarm Module*

The security alarm module is important for the device safety. If the key temperatures are anomaly, it will trigger the protective actions. TDS temperature sensors are distributed in different positions of the superconducting coils. There are six groups of signals used for threshold evaluation. Five groups are used to get the TDS status signal and one group is used to get the TDS fault signal. The program is shown as Fig.4.

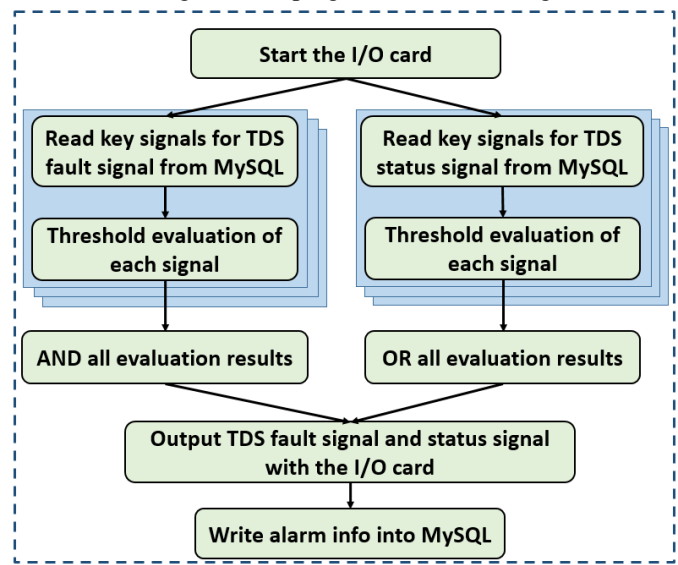

Fig. 4. Security alarm

In the security alarm module, it uses NI PXI-6259 as the I/O card to output the TDS fault signal and status signal. So firstly it starts the I/O card and then read 6 groups of signals from MySQL simultaneously. After threshold evaluation, it AND all evaluation results of fault signal and OR all evaluation results of status signal to get the TDS fault signal and status signal respectively. Finally it outputs the two signals with the I/O card. And the alarm information will be wrote into MySQL.

*D. Information inquiry module*

The information inquiry module provides data services for the TDS administrator to know the TDS running state. It provides three data services: alarm log inquiry, real-time data view, and historical data curve. The administrator can query the latest 20 logs or logs of one day and the relevant alarm logs will be returned from MySQL. When the administrator view the nitrogen data or helium data, it will read all real-time nitrogen signals' data or helium signals' data from MySQL. The MDSPlus shot tree and day tree stored the long-term TDS data. When the administrator requests the data of one or several signals in one shot or several days, it will read historical data from the MDSplus shot tree or day tree, and draws on the interface.

*E. Hardware*

The data communication module and the security alarm are deployed on one device. The nitrogen data acquisition and helium data acquisition are distributed on two devices. The OS of all devices are Windows 7 professional. Table I is a list of the hardware of TDS_DAQ.

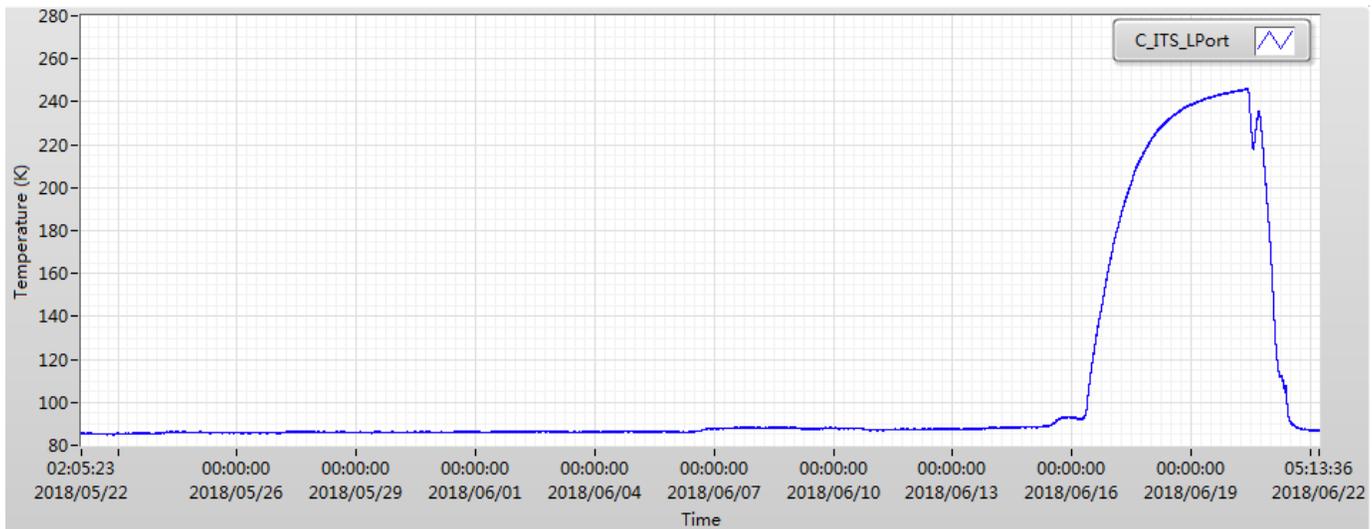

Fig. 5. C_ITS_LPort which is a nitrogen temperature signal in one month

TABLE I
THE HARDWARE OF TDS_DAQ

| Classification | Hardware |
| --- | --- |
| The nitrogen data acquisition | Chassis: NI PXIe-1065<br>Controller: NI PXIe-8840<br>Card: NI PXI-6259 |
| The helium data acquisition | Chassis: ADLINK cPCIS-2630<br>Controller: ADLINK cPS-H325 |
| The security alarm | Chassis: NI PXIe-1082<br>Controller: NI PXIe-8840<br>Card: NI PXI-6259 |

## IV. RESULTS

The TDS_DAQ has been deployed in the winter of 2017. In the 2018 EAST campaign, the TDS_DAQ has continuously run several months. It can long time acquire the TDS data stably and operates totally automatically. The Fig.5 shows the curve of C_ITS_LPort which is a nitrogen temperature signal acquired by the TDS_DAQ in one month.

Furthermore, another primary task of TDS_DAQ is the security alarm. The security alarm module also runs from the very start of the 2018 EAST campaign. It behaved stable and reliable.

## V. CONCLUSIONS

The TDS_DAQ is in charge of continuous data acquisition of the nitrogen and helium temperature signals, TDS security alarm and long-term data storage. It supports long-pulse and continuous acquisition of the nitrogen and helium temperature signals which are important for the administrator to know the state of EAST superconducting coils. It provides the long-term data storage and the administrator can get the TDS information with the information inquiry service or other tools. If the temperature is anomaly, it will trigger safety interlock system to take action to prevent the damage of the EAST device.

TDS_DAQ has been deployed and used in the 2018 EAST campaign. The system has run non-stop and stable about three months, with reaching the designed requirements. Meanwhile, the automatic operation can free the TDS administrator from the tired work. In conclusion, TDS_DAQ reaches the goal of acquiring and monitoring the temperature signals. In the future, we may provide more function and service, involving more data processing and other engineering system. The implementation of TDS_DAQ is also a preliminary study of CFETR (China Fusion Engineering Test Reactor) [12], which will be a new tokamak device of China.

ACKNOWLEDGMENT

The authors would like to thank Weibin Xi, Li Qian, Yun Sha, Huanyu Chen and Hao Wu from Institute of Plasma Physics, Chinese Academy of Sciences for their support during the system test. This work is supported by National Key R&D Program of China (Grant No: 2017YFE0300500, 2017YFE0300504).